# LUMINESCENCE IN ANION-DEFICIENT HAFNIA NANOTUBES

A.O. Shilov[1], R.V. Kamalov[1], M.S. Karabanalov[1], A.V. Chukin[1], A.S. Vokhmintsev[1],
G.B. Mikhalevsky[2], D.A. Zamyatin[1,2], A.M.A. Henaish[1,3], I.A. Weinstein[1,4,*]

[1] NANOTECH Centre, Ural Federal University, Mira street, 19, Yekaterinburg, Russia
[2] Institute of Geology and Geochemistry, Ural Branch of the Russian Academy of Sciences, Vonsovskogo street, 15, Yekaterinburg, Russia
[3] Physics Department, Faculty of Science, Tanta University, Tanta, Egypt
[4] Institute of Metallurgy, Ural Branch of the Russian Academy of Sciences, Amundsena street, 101, Yekaterinburg, Russia

*corresponding author: i.a.weinstein@urfu.ru

**Abstract.** Hafnia-based nanostructures and other *high-k* dielectrics are promising wide-gap materials for developing new opto- and nanoelectronics devices. They possess a unique combination of physical and chemical properties such as insensitivity to electrical and optical degradation, radiation damage stability, a high specific surface area, and an increased concentration of the appropriate active electron-hole centers. The present paper aims to investigate the structural, optical, and luminescent properties of anodized non-stoichiometric $HfO_2$ nanotubes. As-grown amorphous hafnia nanotubes and nanotubes annealed at 700°C with a monoclinic crystal lattice served as samples. It has been shown that the bandgap $E_g$ for direct allowed transitions amounts to 5.65 ± 0.05 eV for amorphous and 5.51 ± 0.05 eV for monoclinic nanotubes. For the first time, we have studied the features of the intrinsic cathodoluminescence and photoluminescence of the obtained nanotubular $HfO_2$ structures with an atomic deficiency in the anion sublattice at temperatures of 10 and 300 K. A broad emission band with a maximum of 2.3 – 2.4 eV has been revealed. We have also conducted an analysis of the kinetic dependencies of the observed photoluminescence for synthesized $HfO_2$ samples in the millisecond range at room temperature. It showed that there are several types of optically active capture and emission centers based on vacancy states in the $O_{3f}$ and $O_{4f}$ positions with different coordination numbers and a varied number of localized charge carriers ($V^0$, $V^-$, and $V^{2-}$). The uncovered regularities can be used to optimize the functional characteristics of developed-surface luminescent media based on nanotubular and nanoporous modifications of hafnia.

## 1. INTRODUCTION

To date, hafnium dioxide is highly sought-after for creating new hardware components in nanoelectronics. Along with the usage of $HfO_2$ as a gate dielectric in CMOS transistors [1], it is beneficial as a functional medium in designing memristor-cell-based memory devices to operate relying on mechanisms involving induced defects of their intrinsic nature [2–5]. It is known that hafnia includes oxygen vacancies of various configurations and charge states as active electron-optical centers [6]. It is these vacancies that localize charges; thereby, $HfO_2$ is able to exhibit its own luminescence and typical memristive behavior [2,3,7,8]. In addition, due to the high atomic mass of a cation, hafnia is an excellent solid-state matrix for doping with rare earth ions. This makes it possible to project up-to-date scintillation media and efficient light-emitting devices [7,9,10].

One of the methods for the targeted synthesis of hafnia with variable anionic non-stoichiometry is to develop low-dimensional structures with a high concentration of surface defects as a consequence of morphological features and non-equilibrium growth conditions. For producing $HfO_2$ nanotubes (NT), various physico-chemical methods can be utilized, such as atomic layer deposition into a matrix of nanoporous aluminum oxide [11], a combination of electro-spinning and ion sputtering [12,13], as well as anodic oxidation of metal foil [14,15]. Whatever technique is used, the grown nanotubes are initially amorphous. When treated at high temperature, their atomic structure passes into tetragonal and more stable monoclinic crystalline phases [16]. In some cases, maintaining the amorphous structure is arduous during annealing as a necessary step in the synthesis. In particular, to remove precursor residues, when synthesizing nanotubes by magnetron sputtering of a hafnium target onto organic PVP fibers, heating to 500°C is required. In this case, $HfO_2$ passes into a monoclinic phase [13].

Against other synthesis methods, electrochemical oxidation is a relatively simple method for growing nanotube arrays on a metal substrate. One of the first successful attempts to obtain anodized $HfO_2$ nanotubes is reported in [14], where a mixture of sulfuric acid and sodium fluoride is applied as an electrolyte. Manipulating the morphology of the synthesized oxide layer – solid → porous → nanotubular – can be performed by varying both the electrolyte composition and parameters of electrochemical synthesis, such as a voltage between the electrodes, current flowing through the solution, oxidation time, and temperature [15]. In turn, the geometric parameters and phase composition of NT have a dramatic influence on the features of their electronic subsystem and various structure-sensitive capabilities [17,18].

To complete the role of anionic vacancy centers in the processes of charge redistribution and transfer, it is necessary to dive deeper into the patterns of forming the energy structure of a material with intrinsic defects and complexes based on them. Conducting experimental research on the optical and luminescent properties of $HfO_2$ NT is extremely important for establishing the aspects of the energy spectrum caused by the morphology of the material. Currently, independent investigations of the optical and luminescent properties of bulk and thin-film hafnia samples doped with various ions, mainly with an amorphous and monoclinic structure, are known [9,19–22]. Simultaneously, examining the developed surface structures of hafnia could provide insight into the role of surface optically active centers and defects of various natures in the radiative and non-radiative relaxation processes of electronic excitations in $HfO_2$ nanotubes. The goal of the present work is to analyze the characteristics of the luminescent response of nanotubular structures of hafnia with an atomic deficiency in the anion sublattice.

## 2. EXPERIMENTAL

### 2.1 Samples

In this work we investigate the properties of three samples: nanotubular hafnia (as-grown and annealed NT) and hafnia nanopowder. The nanotubular samples were synthesized using electrochemical anodization of the hafnium foil, where Hf acted as the anode and stainless steel was used as the cathode. The anodization was carried out in the solution containing $NH_4F$ (0.5 wt%), $H_2O$ (2 wt%) and ethylene glycol. The synthesis was performed under the constant voltage of 40 V for 4 hours. The synthesized samples contain carbon and fluorine as a part of the electrolyte used in their synthesis. These precursor residues can be removed after heating hafnia NT to 300 °C, resulting in obtaining as-grown sample. Besides, the NT were annealed in air at 700 °C for 2 hours – annealed sample. Crystalline hafnia nanopowder (HFO-1 grade, TU 48-4-201-72) is used as a reference structure [23].

### 2.2 Experimental techniques

The morphology was investigated using transmission electron microscope (TEM) JEOL JEM-2100 and scanning electron microscope (SEM) Carl Zeiss SIGMA VP with Oxford Instruments X-Max 80 module for energy dispersive analysis (EDS) to determine chemical composition of the sample. X-ray diffraction (XRD) was analyzed using Shimadzu XRD-7000 diffractometer. XRD-patterns were measured for the 2θ range from 10° to 85° with 0.06° step. Raman spectrum was registered by LabRAM HR Evolution spectrometer in the range of 50 – 2000 $cm^{-1}$. Integrated red solid-state laser with wavelength of 633 nm was used as an excitation source. FTIR characterization was performed using Bruker Vertex 70 spectrometer with an integrating sphere attachment. Hafnia samples were mixed with KBr powder as diluent in order to obtain diffuse reflectance spectra in the range 400 – 4000 $cm^{-1}$. The obtained spectra were analyzed using Kubelka-Munk formalism.

Diffuse reflectance spectra (DRS) were measured in the 210-850 nm range with the step of 0.1 nm using two-beam spectrophotometer SHIMADZU UV-2450 and integrating sphere ISR-2200 attachment. Barium sulfate powder was used as a white body reference.

Cathodoluminescence (CL) spectra were registered with a HORIBA H-CLUE spectrometer. Photoluminescence (PL) measurements were taken using an Andor Shamrock SR-303i-B spectrograph with a NewtonEM DU970P-BV-602 CCD recording array. A DTL-389QT ultraviolet laser with a wavelength of 263 nm (4.71 eV) was used as a photoexcitation source. PL excitation spectra and kinetic dependencies were recorded using a Perkin Elmer LS 55 spectrometer equipped with a xenon source. During measurements of the PL decay kinetics, the delay after turning off the Xe lamp was 0.1 ms and increased to 60 ms in steps of 0.1 ms. The signal recording time amounted to 12.5 ms with a total duration of one measurement cycle of 80 ms. In the recording channel, the monochromator slit width was 20 nm, whereas in the excitation channel it was 10 nm. The luminescence was excited by 243 nm (5.1 eV) photons for decay curve registration. To plot the obtained luminescence spectra against photon energy a correction was performed which is necessary in case of using diffraction grating with linear dispersion in wavelengths [24]. Measurements of CL and Raman spectra were carried out at the Common Use Center "Geoanalyst" (Institute of Geology and Geochemistry of the Ural Branch of RAS, Ekaterinburg).

## 3. RESULTS AND DISCUSSION

3.1 Electronic microscopy

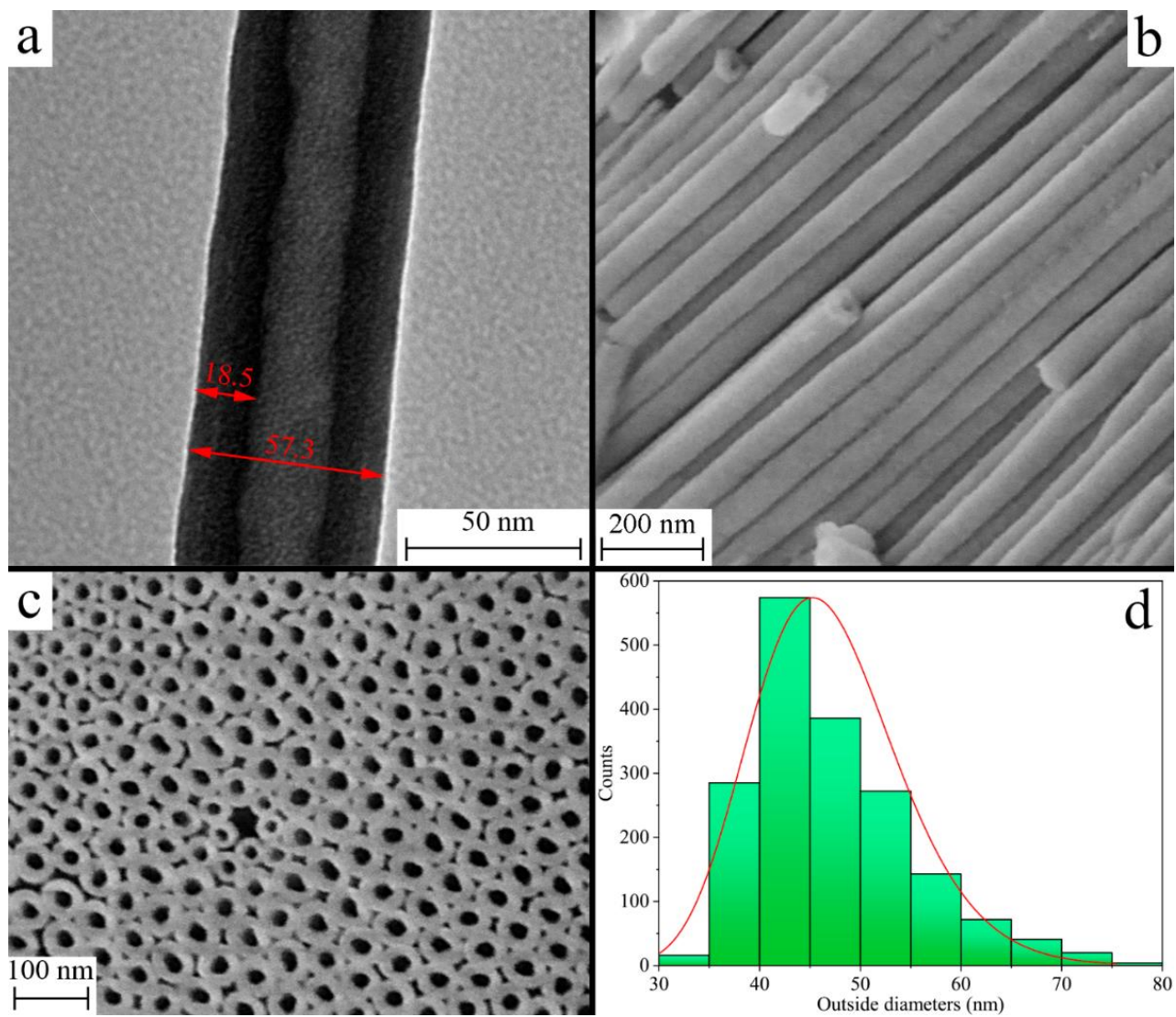

**Figure 1.** *Images of synthesized nanotubes obtained using TEM (1a), SEM (1b, 1c). The size distribution of nanotube diameters is shown in Figure 1d.*

Figure 1 shows SEM and TEM images of the grown hafnia nanotubular arrays. The length and average diameter of the synthesized nanotubes are found to be $10 \pm 3$ μm and $46 \pm 7$ nm, respectively. The diameter distribution depicted in histogram 1d obeys a lognormal law with parameters $\mu = 3.83$ and $\sigma = 0.16$. A chemical analysis conducted by the EDS method detected no impurities of heavy elements. After annealing up to 300 and 700 °C, only hafnium and oxygen are observed in the EDS spectra.

## 3.2 Structural-phase analysis

Figure 2a outlines X-ray diffraction (XRD) analysis data for nanotubular structures. For as-grown NT, a halo is observed in the range of 20–38°C, i.e., it can be argued that these structures are amorphous. In the process, the most intense peaks match a hafnium foil utilized for growing the oxide layer. In addition, there are several peaks indicating small inclusions of cubic and monoclinic phases of $HfO_2$. When annealed, the samples exhibit no halo, and hafnia is in the monoclinic phase. For comparison, Figure 2a shows the diffraction pattern of monoclinic $HfO_2$ nanopowder.

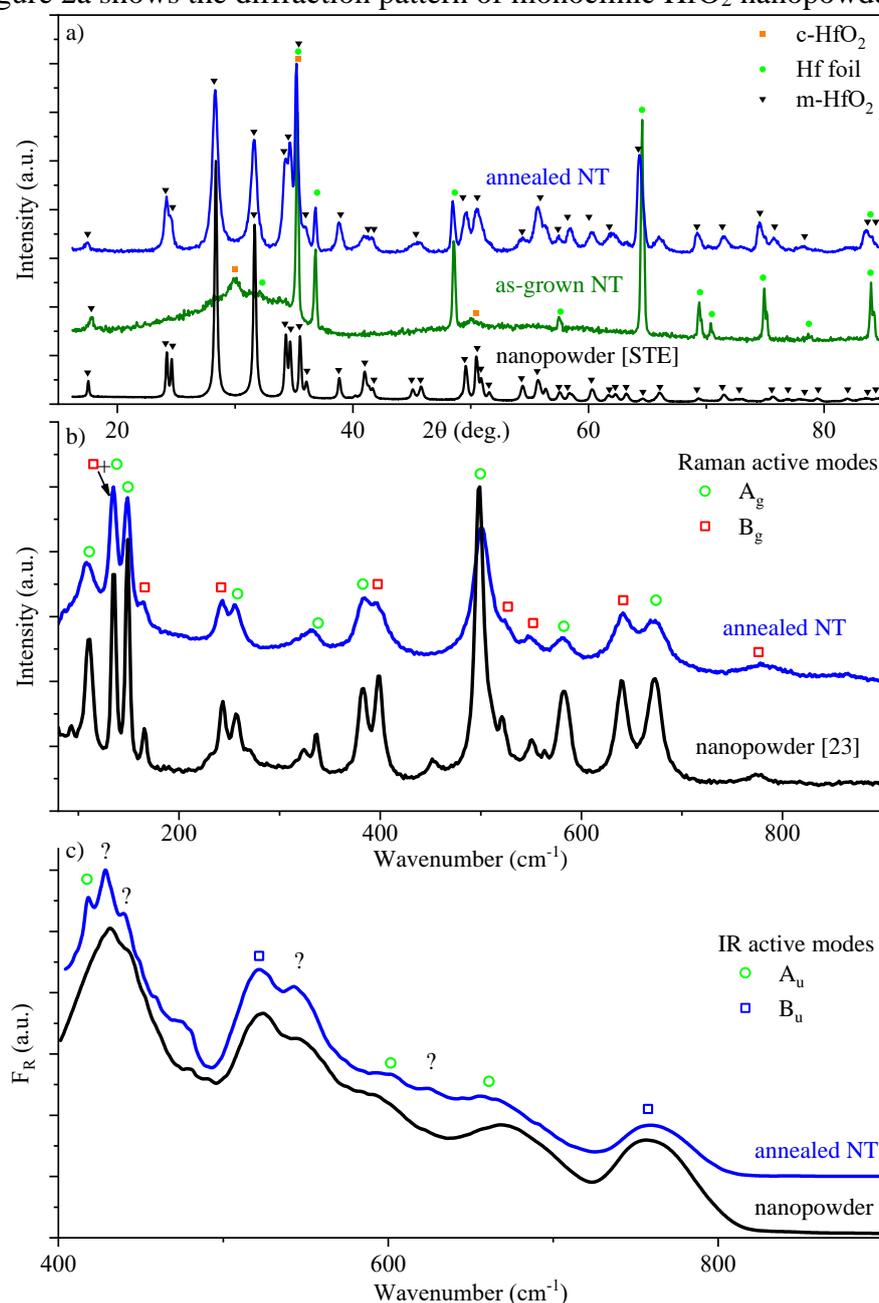

**Figure 2.** *Results of XRD analysis (a), Raman (b) and IR (c) spectroscopy for synthesized $HfO_2$ nanotubes in comparison with data for $HfO_2$ nanopowder with a monoclinic structure*

Raman and IR spectroscopy confirm the structural transition to a monoclinic crystalline phase, observed for the nanotubes after annealing – the data are presented in Figures 2b and 2c, respectively. The Raman spectrum revealed seventeen active vibrational $A_g$ and $B_g$ modes. Within the present research, the IR spectrum of hafnia nanotubes was measured for the first time, as shown in Figure 2c. Table 1 lists the phonon energies for IR-active vibrational modes in comparison with known data for thin films [25–27] and nanoparticles synthesized by precipitation [28] and the sol-gel method [29]. An analysis of the measured IR spectrum using the predicted data in [30] made it possible to identify five modes; see Figure 2c and Table 1. In addition, there are maxima experimentally held fixed both in the nanotubes and in the monoclinic hafnia nanopowder, however, the group-theoretical analysis [28] did not determine their type (see Table 1). For as-grown nanotubes with an amorphous structure, the Raman and IR spectra contained no active vibrational modes.

**Table 1.** Phonon energies of IR-active vibrational modes for the hafnia nanostructures studied

| Peak energy, cm$^{-1}$ | | | |
|---|---|---|---|
| This work | | Independent experiment | Calculation [30] |
| Annealed NT | Monoclinic nanopowder | | |
| 417 | — | 410 [25,26]<br>415 [28] | 410, $A_u$ |
| 428 | 432 | — | ? |
| 440 | 443 | — | ? |
| 523 | 523 | 506 [27]<br>515 [26]<br>516 [29] | 512, $B_u$ |
| 543 | 546 | 550 [27]<br>555 [29] | ? |
| 602 | 592 | 595 [27]<br>600 [25]<br>615 [26] | 600, $A_u$ |
| 626 | – | 625 [26]<br>635 [25] | ? |
| 666 | 669 | 680 [29] | 665, $A_u$ |
| 759 | 759 | 740 [26]<br>750 [29]<br>752 [25] | 730, $B_u$ |

## 3.3 Estimation of the bandgap width

Using the Kubelka-Munk formalism [23,31], we analyzed diffuse reflectance spectra recorded at room temperature (experimental data is depicted in Figure 3a). The resulting absorption spectrum $F_R(h\nu)$ (see Figure 3b) includes a broad Gaussian-shaped shoulder that emerged because of oxygen vacancies: their absorption lies within energy region of $h\nu > 3$ eV [32].

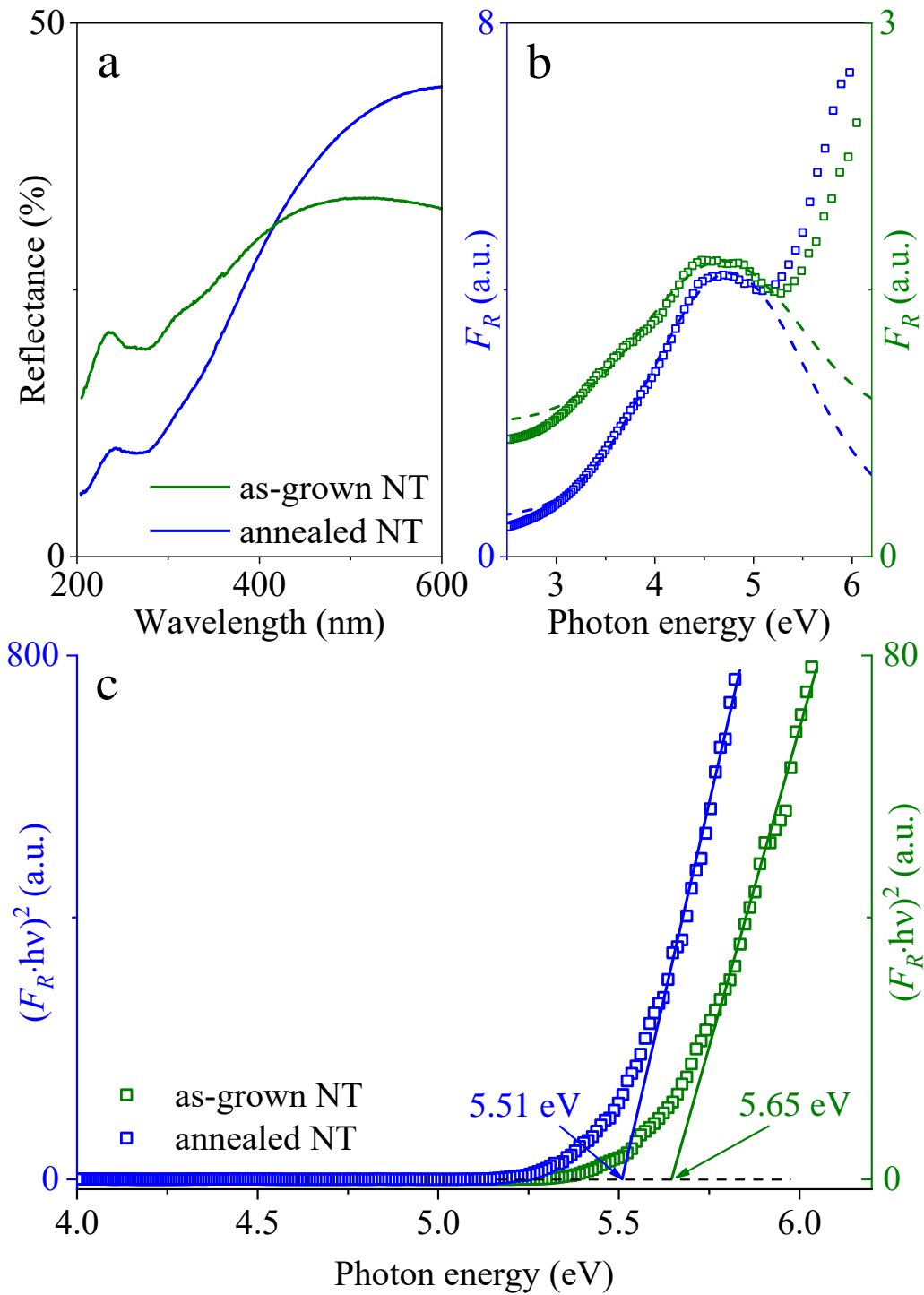

**Figure 3.** *(a) The diffuse reflectance spectra measured for as-grown (green) and annealed (blue) nantotubes; (b) Absorption spectra $F_R(hv)$ converted using Kubelka-Munk formalism, dashed curves denote a Gaussian-shaped shoulder that emerged because of oxygen vacancies' absorption; (c) The intrinsic absorption edge in Tauc coordinates for direct allowed transitions taking defects' absorption into account.*

In order to correctly estimate the bandgap width using the Tauc approach, it is necessary to extract the optical absorption caused by active defects within region of the spectral shoulder. The latter was approximated by a Gaussian (see the dashed curves in Figure 3b) and subtracted from the optical absorption spectra. Figure 3c includes revised curves $F_R(hv)$ plotted in Tauc coordinates for direct allowed band-to-band transitions [33]. In frame of the approach used, the optical gap width is $E_g = 5.65 \pm 0.05$ eV for as-grown and $5.51 \pm 0.05$ eV for annealed nanotubes. It is worth stressing that the above evaluations were pioneered for nanotubular $HfO_2$ and that no independent data were found to allow for comparison. Similar redshift of the bandgap was previously observed for hafnia thin films in [34]. The authors claim that the indirect bandgap in amorphous as-grown films is 5.60 eV, which decreases as the annealing temperature grows, reaching 5.13 eV for thin films annealed at 600°C with a monoclinic crystal structure. Nevertheless, the values obtained are reconciled with previous estimates of direct $E_g$ for amorphous and monoclinic hafnia in various structural modifications, such as nanopowder (in range of 5.5-5.8 eV) [23,35–37], thin films (5.3–5.8 eV) [38–40], bulk single crystal (5.89 eV) [41], etc.

In hafnia, we have previously detected an intrinsic absorption edge formed through direct and indirect allowed band-to-band transitions [23,36]. However, due to the distortion of the intrinsic absorption edge in the low-energy region, we did not succeed in determining indirect optical gap width for as-grown and annealed NT.

### 3.4 Spectral characteristics of luminescence

When measuring photo- and cathodoluminescence in the samples under study, a blue-green emission is observed at different temperatures (see Figure 4). It is characteristic of hafnia in various morphological modifications such as thin films [6,21,39,42], nanopowders [23,43,44], nanocrystals [45,46], and bulk single crystals [41]. It should be noted that cooling the nanotubes to a temperature of 10 K has almost no influence on the broad Gaussian-shaped bands in the PL spectra (see inset in Figure 4). It is known that the observed emission is associated with the presence of oxygen vacancies in the hafnia [6,32,43,44]. Table 2 provides the values of the maximum energies and half-widths calculated during the approximation of the luminescence spectra by Gaussians, as well as the estimates of the intensity ratio at different temperatures.

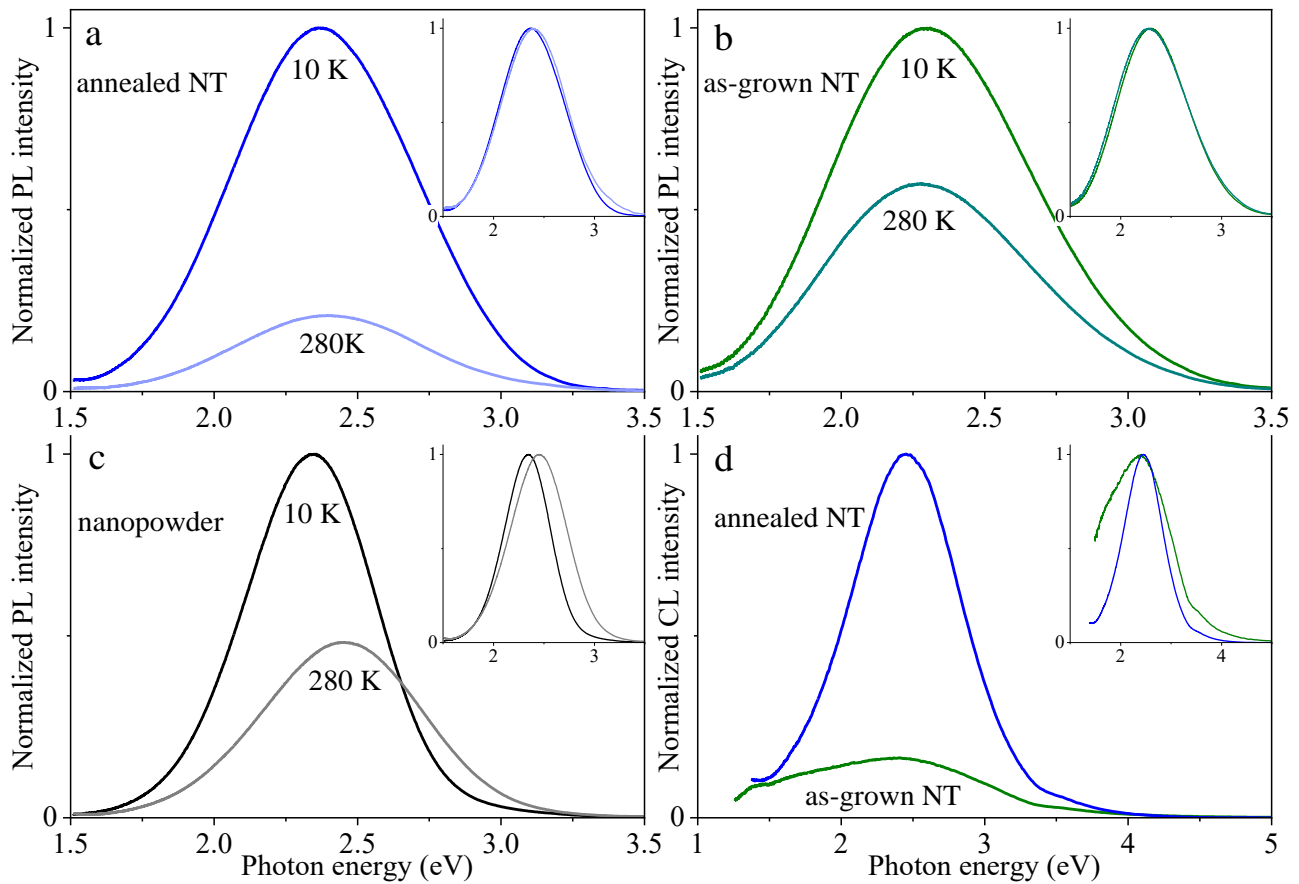

**Figure 4.** *PL spectra for the studied NT at temperatures of 10 K and 280 K (a, b). Spectra for as-grown NT are shown in green, and spectra for NT after annealing are shown in blue. For comparison, the PL spectrum of the nanopowder is shown (c). Cathodoluminescence spectra (d) were measured at room temperature. The spectra normalized to their own maxima for a visual shape comparison are depicted in the insets.*

The photoluminescence excitation spectra (PLE) of hafnia nanotubes can be seen in Figure 5. They contain four maxima, and their positions are almost unchanged after high-temperature treatment. The peaks at 4.6, 4.8 and 5.1 eV fall into the optical absorption region for oxygen vacancies [32] (see also the spectral shoulder in Figure 3b), and the 5.5 eV maximum rests in the region of optical band-to-band transitions near the intrinsic absorption edge (see Figure 3). The comparison of normalized PL spectra in Figure 5 points to the fact that the luminescence band for as-grown NT at room temperature is slightly wider and insignificantly shifted to lower energies in comparison with a similar peak for annealed NT.

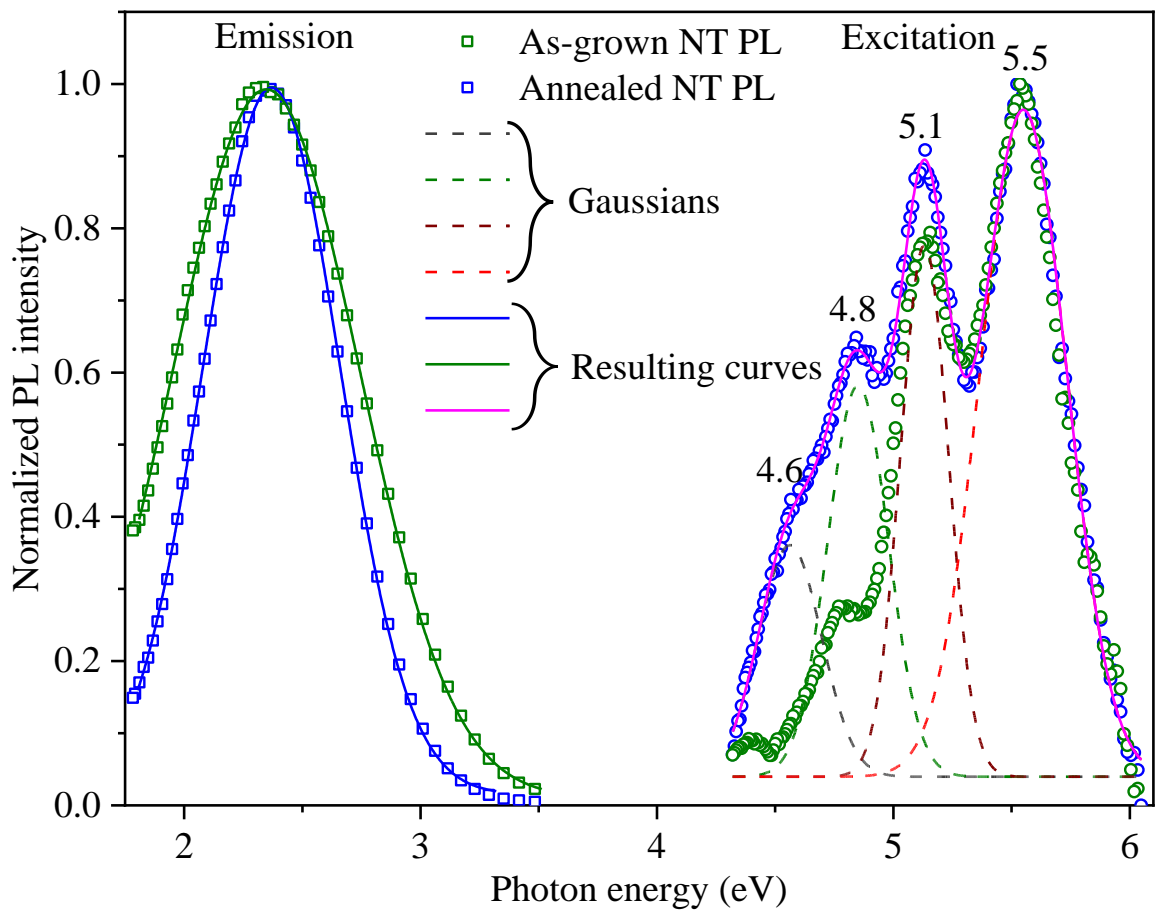

**Figure 5.** *PL emission (squares) and excitation (circles) spectra of as-grown (green) and annealed (blue) NT. The solid and dashed lines indicate an approximation of the spectra by Gaussian curves and individual components, respectively.*

**Table 2.** Energy parameters for the studied luminescence spectra

| Sample, excitation wavelength | Temperature | $E_{max}$, ± 0.02 eV | FWHM, ± 0.02 eV | $I_{10} / I_{280}$ | Reference |
|---|---|---|---|---|---|
| | | Photoluminescence | | | |
| As-grown nanotubes, 263 nm | 10 | 2.30 | 0.82 | 1.75 | This work |
| | 280 | 2.27 | 0.84 | | |
| Annealed nanotubes, 263 nm | 10 | 2.38 | 0.74 | 4.78 | This work |
| | 280 | 2.40 | 0.75 | | |
| Monoclinic powder, 263 nm | 10 | 2.33 | 0.54 | 2.07 | This work |
| | 280 | 2.45 | 0.66 | | |
| Monoclinic nanotubes, 325 nm | RT | 2.91* | 0.98* | — | [12] |
| Monoclinic nanopowder, 210 nm | RT | 2.60 | 0.58 | — | [23] |
| Monoclinic nanopowder, 200 nm | 10 | 2.96* | 0.91* | 1.05* | [43] |
| | RT | 2.84* | 0.76* | | |
| | | Cathodoluminescence | | | |
| As-grown nanotubes | RT | 2.40 | 1.32 | — | This work |
| Annealed nanotubes | RT | 2.45 | 0.93 | — | This work |
| Monoclinic nanopowder | RT | 3.00 | 1.04 | — | [43] |
| Amorphous films | RT | 2.75 | 0.87 | — | [42] |

*Our estimates based on the data given in the cited paper

The papers [12,13] investigated the photoluminescence of HfO$_2$ nanotubes produced by magnetron sputtering of a hafnium target onto polyvinylpyrrolidone nanofibers doped with rare earth ions. According to [12], upon excitation by laser irradiation at 325 nm, emission with a maximum of 2.9 eV is observed. It is blue-shifted relative to our data obtained with excitation at 263 nm, see Table 2. The shift of the emission maximum may be associated with different PL excitation energies, as well as with different methods of synthesis of the nanostructures at hand. In particular, the average diameter of nanotubues in [12,13] is 200 nm, which is higher by a factor 4 than the similar magnitude for the NT grown by anodic oxidation in our work.

The authors of the present paper were the first to measure the cathodoluminescence (CL) spectra of hafnia nanotubes, see Figure 4d and Table 2. It can be seen that the positions of the maxima in the PL and CL spectra virtually coincide. In addition, when subjected to high-temperature annealing, the samples exhibit a narrowing and a slight shift of the emission bands towards the short-wavelength region of the spectrum. Moreover, in contrast to the PL response, the CL spectrum for amorphous nanotubes has a more complicated shape that cannot be approximated by a single Gaussian. Since there are no independent experimental data for nanotubes, we can compare only our results with the CL spectra for monoclinic nanopowder and thin films of non-stoichiometric-in-oxygen sublattice hafnia [42,43]. In the first case, the CL band shifts towards the high-energy region [43], and its half-width is quite consistent with our data. The CL spectrum for amorphous thin films is a superposition of two Gaussian components with maxima near 2.0 and 2.6 eV [42]. The authors also report shift of the emission maximum in the blue range from 2.6 to 2.75 eV, associating it with the different concentration of OH-groups in the films synthesized using of two precursor systems in atomic layer deposition.

3.5 Photoluminescence decay

In order to study the kinetic peculiarities of the observed PL, a quantitative analysis of the decay curves was carried out. Emission was recorded at spectral maxima for each sample, such as 2.27 eV (as-grown NT), 2.4 eV (annealed NT), and 2.45 eV (nanopowder). Figure 6 shows the measured time dependencies.

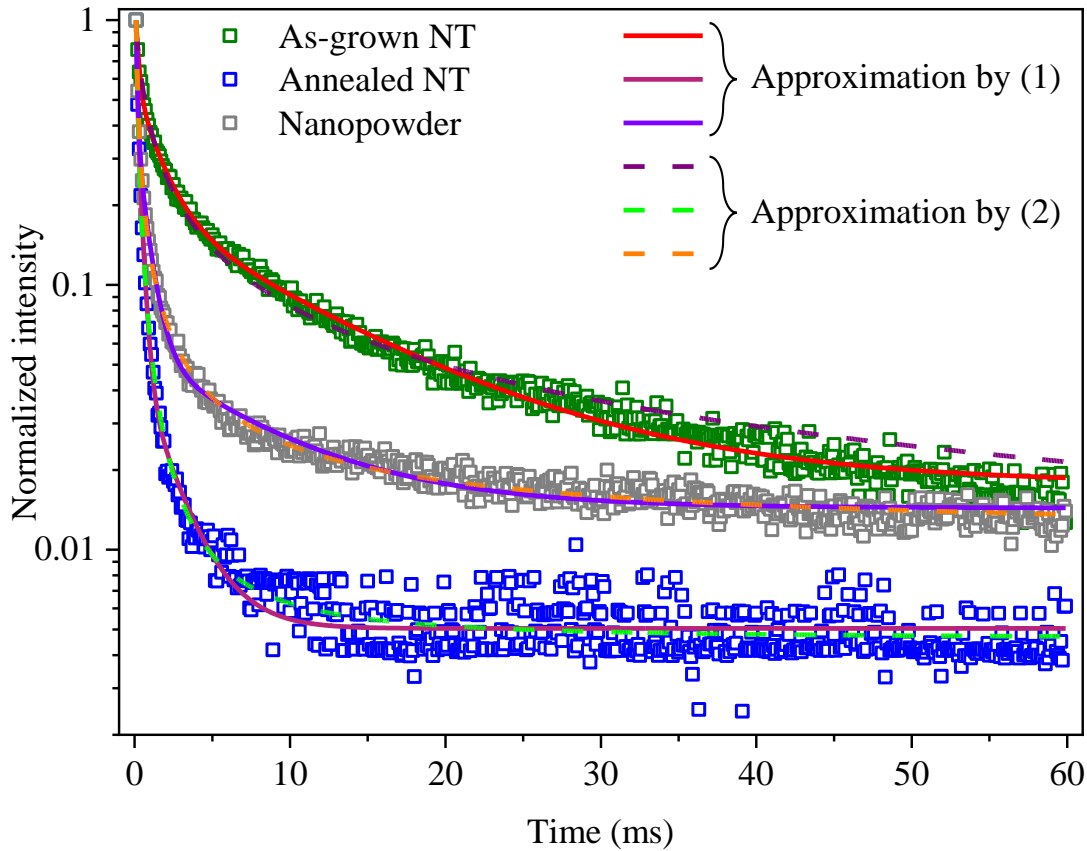

**Figure 6.** *Photoluminescence decay curves in HfO$_2$ nanotubes and nanopowder. Experimental data (squares) are approximated using expressions (1) and (2); the approximation curves are shown by solid and dashed lines, respectively.*

The observed PL decay can be described as a superposition of three exponential components:

$$I(t) = \sum_{i=1}^{3} A_i \cdot e^{-t/\tau_i}, \qquad (1)$$

where $A_i$ are constants; $\tau_i$ is decay times, ms. It should be also underscored that the same experimental results can be satisfactorily approximated using Becquerel's empirical formula:

$$I(t) = \frac{I_0}{(1 + t/C)^r}, \qquad (2)$$

where *C* is a constant, ms; *r* is an index whose value is related to the ratio of the effective cross-sections of traps and emission centers [47,48].

The parameters of approximation using the expressions (1) and (2) are presented in Table 3. Similar PL decay times for hafnia nanopowder annealed at 450 °C were obtained in [45]. According to the latter, such temperature provokes a noticeable growth of crystalline domains, i.e., inclusions of the amorphous phase may explain the similarity of decay times.

**Table 3.** Kinetic parameters of photoluminescence in $HfO_2$ nanosructures

| Sample | Multiexponential decay, see (1) | | | | | |
|---|---|---|---|---|---|---|
| | $A_1$ | $A_2$ | $A_3$ | $\tau_1$, ms | $\tau_2$, ms | $\tau_3$, ms |
| as-grown NT | 0.744 | 0.382 | 0.176 | 0.191 | 1.56 | 11.5 |
| annealed NT | 2.595 | 0.726 | 0.048 | 0.056 | 0.29 | 2.12 |
| nanopowder | 1.777 | 0.360 | 0.042 | 0.096 | 0.72 | 7.92 |
| nanopowder [45] | 0.284 | 0.406 | 0.306 | 0.198 | 2.04 | 11.5 |

| Sample | Becquerel's decay, see (2) | |
|---|---|---|
| | $C$, ms | $r$ |
| as-grown NT | 0.357 | 0.739 |
| annealed NT | 0.181 | 1.591 |
| nanopowder | 0.111 | 0.950 |

The revealed dependencies indicate the non-elementary nature of relaxation and charge transfer processes. In particular, hafnia's oxygen vacancies in various configurational and charge states can participate in luminescence mechanisms. The higher the index *r*, estimated for annealed nanotubes, the lower the ratio of the effective cross-sections of traps and emission centers. In other words, the index *r* increases with a decrease in the concentration of defects acting as charge traps at room temperature [47,48]. This gives rise to a faster decay of the PL of annealed NT in comparison with that of the as-grown nanotubular structures and nanopowder. This is a consequence of annealing in an air atmosphere, due to which the number of oxygen vacancies should diminish.

It is known that there are two types of oxygen vacancies in monoclinic $HfO_2$, namely, $O_{3f}$ (with coordination number 3) and $O_{4f}$ (with coordination number 4), which can also be in different charge states [32,49,50]. According to DFT calculations [32,46], the $O_{4f}$-vacancies are more stable compared to those of $O_{3f}$. In turn, among $O_{4f}$, vacancies $V^0$, $V^-$, and $V^2$ with two, three, and four electrons, respectively, are the most stable; whereas among $O_{3f}$, fully ionized $V^{2+}$- and $V^+$-vacancies are faced [32,50]. These defects originate energy levels in the bandgap, as well as near the bottom of the conduction band of hafnia. This fact explains the emergence of a shoulder in the optical absorption spectrum for hν > 3 eV. In the PLE spectra, four Gaussian components of 4.6, 4.8, 5.1, and 5.5 eV indicate that photoluminescence is effectively excited by optical transitions to levels located near the bottom of the conduction band; the former may be unoccupied levels of $V^-$, $V^+$ and $V^{2+}$, correspondingly [50] and the latter may correspond to band-to-band optical transitions.

## 4. CONCLUSION

In this work, we studied the luminescence features of hafnia nanotubes synthesized by electrochemical oxidation in a potentiostatic regime. It was shown that the nanotubes were grown with a length of 10 ± 3 μm and an average outer diameter of 46 ± 7 nm. According to XRD analysis, Raman and IR spectroscopy, the as-grown NT had an amorphous structure, further annealing at a temperature of 700°C led to crystallization and their transition to a monoclinic phase.

We measured diffuse reflectance spectra at room temperature, followed by their analysis within the Kubelka-Munk formalism. It was established that a spectral shoulder caused by optically active anion vacancy centers appears in the energy range of hν > 3 eV and distorts the intrinsic absorption edge. Within the modified Tauc approach, we gained estimates of the optical gap width formed by direct allowed band-to-band transitions: $E_g = 5.65 \pm 0.05$ eV and $5.51 \pm 0.05$ eV for as-grown and annealed nanotubes, respectively. The findings secured correspond to similar characteristics for hafnia in various morphological modifications.

We measured for the first time cathodoluminescence spectra for $HfO_2$ NT of various phase compositions and also conducted a study of their photoluminescent properties in the temperature range of 10 – 300 K. It was shown that cooling to 10 K has virtually no influence on the shape of the emission band for the synthesized nanotubes, the maximum of which is in the region of 2.3 – 2.4 eV. The observed PL and CL are due to processes involving vacancy-based centers in the anionic sublattice of hafnia in various configurational ($O_{3f}$ and $O_{4f}$) and charge states ($V^-$, $V^+$ and $V^{2+}$).

The PL kinetic curves handle several types of capture centers in nanotubular hafnia. Upon high-temperature annealing up to 700°C, their concentration diminishes, and a crystalline phase with monoclinic symmetry forms. The revealed regularities may be of practical value in optimizing emission properties for developing luminescent media based on hafnia nanotubes of various phase composition.

## ACKNOWLEDGMENTS

This work was supported by the Russian Science Foundation under grant no. 23-22-00310, https://rscf.ru/project/23-22-00310/.